\documentstyle[11pt,newpasp,epsfig]{article}
\begin{document}
\title{Indicator of Exo-Solar Planet(s) in the Circumstellar Disk
Around $\beta$ Pictoris}

\author{Nick Gorkavyi}
\affil{NRC/NAS; Code 685, NASA/GSFC, Greenbelt, MD, 20771, US}
\author{Sara Heap}
\affil{Code 681, NASA/GSFC, Greenbelt, MD, 20771, US}
\author{Leonid Ozernoy}
\affil{5C3, George Mason University, Fairfax, VA 22030-4444, US}
\author{Tanya Taidakova}
\affil{Computational Consulting Service, MD, 20740, US}
\author{John Mather}
\affil{Code 685, NASA/GSFC, Greenbelt, MD, 20771, US}

\begin{abstract}
Our efficient numerical approach has been applied to modeling the
asymmetric circumstellar dust disk around $\beta$ Pictoris as 
observed with the HST/STIS. We present a new model on the origin of the
warping of the $\beta$ Pic disk. We suggest that the observed warp is
formed by the gravitational influence of a planet with a mass
of about 10 masses of Earth, at a distance of 70 AU, and a small
inclination ($\sim 2.5^\circ$) of the planetary orbit to the main
dust disk. 
Results of our modeling are compared with the STIS observations.
\end{abstract}

\section{Introduction}
In circumstellar disks, the major sources of dust are thought to be 
minor bodies, like comets. Both the stellar radiation drag 
(Poynting-Robertson drag) and stellar wind drag tend to induce 
dust inflow toward the star. As the dust passes by planets 
in its infall, it interacts with them, for example, by accumulating 
in planetary resonances (Liou \& Zook 1999, Ozernoy et al. 2000). 
     The inner region of the $\beta$ Pictoris dusty disk is tilted by a
few degrees with respect to the outer disk (Burrows et al.
1996, Mouillet et al. 1997, Heap et al. 2000 (for references, 
see the last paper). Heap et al. (2000) concluded that the size 
and shape of the warp favors the presence of a planet(s) in 
a slightly inclined orbit.

\section{Modelling $\beta$ Pictoris Disk with Embedded Planet}

The dynamics of interplanetary particles are determined by several 
effects, including stellar radiation pressure; the Poynting-Robertson 
(P-R) and stellar wind drags; resonance effects by planets; and
gravitational encounters with planets. 
 
     To simulate the gravitational dynamics of comets and
nonconservative dynamics of dust particles around a star, we use an
implicit second-order integrator (Taidakova \& Gorkavyi, 1999). 
     In the absence of a planet, the surface density of dust between the 
star and the pericenter of dust sources is constant, while the number density
of dust varies as $r^{-1}$, $r$ being heliocentric distance (Gorkavyi et al.
1997). The presence of a planet induces dramatic changes in the cometary and 
dust distribution. This is illustrated in Fig.~1, which shows 
a representative trajectory of dust particles.

\begin{figure} [!ht]
\centerline{\epsfig{file=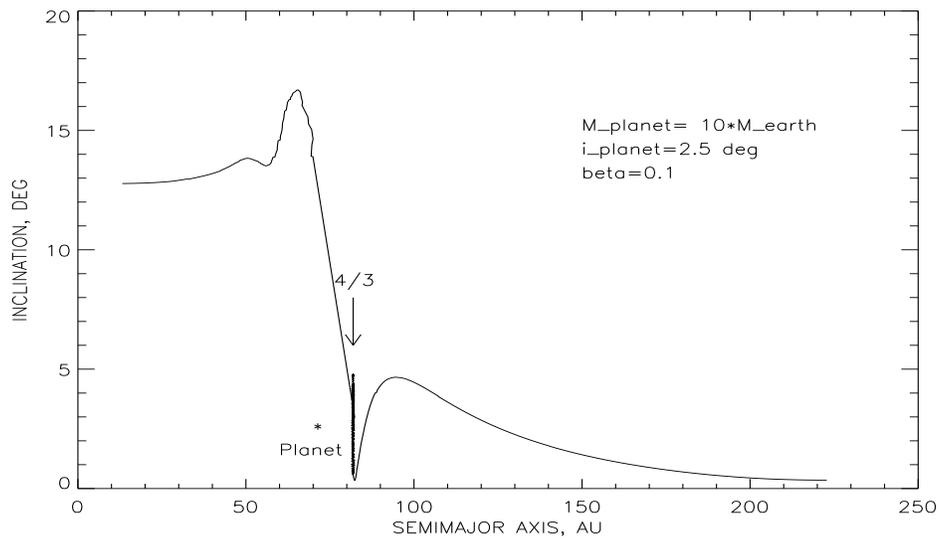,width=5.25in,height=3in}}
\caption{
Representative trajectory of a dust particle
($\beta=0.1$) in the disk of $\beta$ Pic.
P-R drag and a slow precession of the particle's orbit around the orbital
plane of the planet ($a$=70 AU; $i=2.5^\circ$) are the dominating dynamical
processes during its initial inflow (220-83 AU).
The particle's inclination oscillates from 0 to 2$i_{planet}=5^\circ$.
The vertical line marks the resonant capture into the 4:3 resonance.
The next `jump' to $\sim 17^{\circ}$ is due to scattering of the particle
by the planet. The final part of trajectory ($<60 AU$) marks the
particle's drift. 
After strong interactions with the planet, dust particles (and comets) moves
 symmetrically about the new plane of the planetary orbit.  
}
\label{fig1}
\end{figure}

\begin{figure} [!ht]
\centerline{\epsfig{file=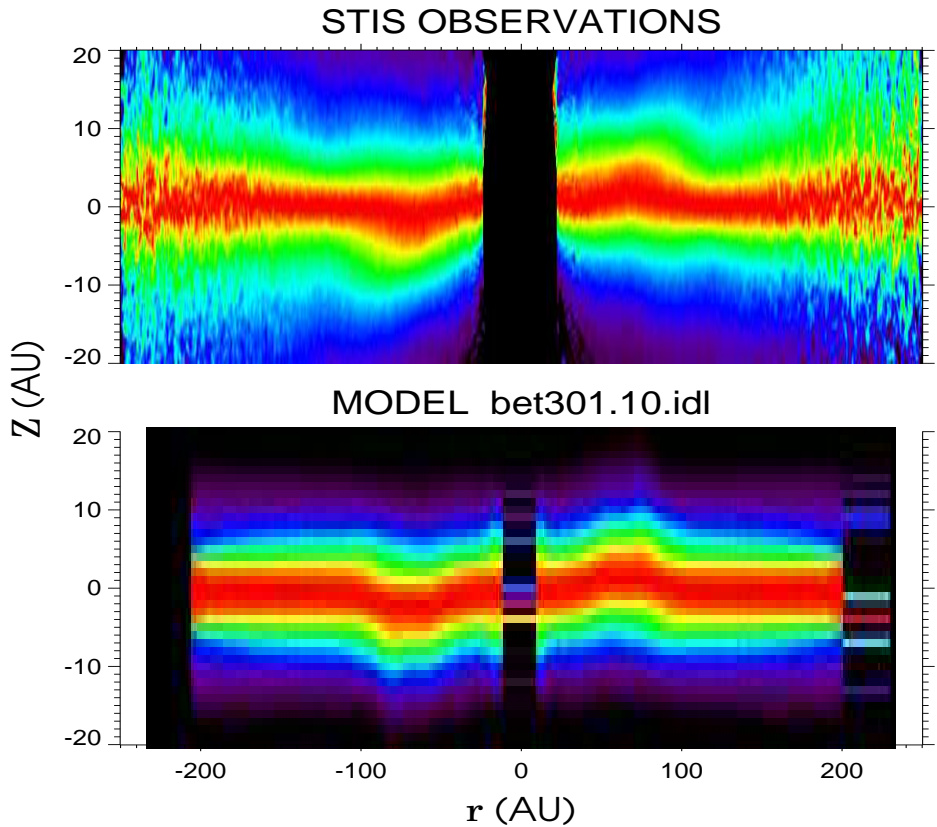,width=5.25in,height=4.44in}}
\caption{
     Warp of the inner part of the $\beta$ Pictoris disk. Upper
panel: STIS/CCD coronagraphic images 
of the $\beta$ Pictoris disk normalized to the maximum flux with the
vertical scale expanded by 4X (Heap et al. 2000).
Bottom panel: 
model disk with $M_{pl}=10M_{\earth}$, orbital radius $r$=70
AU, and inclination $i=2.5^\circ$. The lifetime of 
dust particles due to collisions could be as small as 100 years. 
The stellar radiation pressure is negligible (dust particles
are large enough). This warp is stationary, in contrast to the 
model of Larwood and Papaloizou (1997).
}
\label{fig2}
\end{figure}

\newpage

Two algorithms for modeling the $\beta$ Pictoris disk were 
considered.

{\it Algorithm I}:

1. Input an initial symmetric cometary disk.

2. Input the mass and orbital radius of the planet on an orbit inclined  
   to the initial cometary disk.

3. Simulate the asymmetric dust distribution on a time scale of 
   $1.4\times 10^6$ yrs.
   Inclination of particles orbits to the plane of the initial cometary disk 
   and asymmetry of dust is the result of precession of drifting particles 
   around the planet's orbital plane (see Fig.~1).

4. Simulate the dust-scattered light distribution.

\newpage
{\it Algorithm II}:

1. Input an initial symmetric cometary disk.

2. Input the mass and orbital radius of the planet on an orbit inclined   
   to the initial cometary disk.

3. Determine  the asymmetric distribution of comets due to 
   orbital precession and gravitational scattering by the planet (these 
   processes can be realized without any drifts) after $2.8\times 10^6$ yrs.

4. Simulation of the asymmetric dust distribution for short life 
   times of particles. Asymmetry of dust is mainly a result of an asymmetry 
   in the initial cometary distribution.

5. Simulate the dust-scattered light distribution (see Fig.~2).

\medskip
\noindent
Both alternatives produce a similar warp, but the second alternative seems
to be more preferable due to:
\begin{itemize}
\item a more realistic (shorter) lifetime of particles;  
\item  accounting for gravitational interactions between comets 
and the planet.
\end{itemize}
\noindent 
The STIS images show the disk to be warped at small distances
to the star in a sense that the inner disk is tilted
by $4.6^\circ$  with respect to the outer disk (Heap et al. 2000).
See Fig.~2 to compare the STIS observations with our modeling.

\section{Conclusions}

Asymmetric structures in the circumstellar
disk indicate at least one planet 
embedded in the disk of $\beta$ Pictoris.

\medskip
{\it Acknowledgements}. N.N.G. has been supported through NAS/NRC
Associateship Research program and the STIS Science Team at Goddard. 
NASA grant NAG-7065 to George Mason University
is gratefully acknowledged.

\end{document}